# Word differences in news media of lower and higher peace countries revealed by natural language processing and machine learning

Larry S. Liebovitch* (Queens College, City University of New York and Columbia University, USA)
William Powers (Queens College, City University of New York, USA)
Lin Shi (Queens College, City University of New York, USA)
Allegra Chen-Carrel (University of San Francisco, USA)
Philippe Loustaunau (Vista Consulting LLC, USA)
Peter T. Coleman (Columbia University, USA)

*corresponding author: email: LSL2140@columbia.edu

## Abstract

Language is both a cause and a consequence of the social processes that lead to conflict or peace. “Hate speech” can mobilize violence and destruction. What are the characteristics of “peace speech” that reflect and support the social processes that maintain peace? This study used existing peace indices, machine learning, and on-line, news media sources to identify the words most associated with lower-peace versus higher-peace countries. As each peace index measures different social properties, there is little consensus on the numerical values of these indices. There is however greater consensus with these indices for the countries that are at the extremes of lower-peace and higher-peace. Therefore, a data driven approach was used to find the words most important in distinguishing lower-peace and higher-peace countries. Rather than assuming a theoretical framework that predicts which words are more likely in lower-peace and higher-peace countries, and then searching for those words in news media, in this study, natural language processing and machine learning were used to identify the words that most accurately classified a country as lower-peace or higher-peace. Once the machine learning model was trained on the word frequencies from the extreme lower-peace and higher-peace countries, that model was also used to compute a quantitative peace index for these and other intermediate-peace countries. The model successfully yielded a quantitative peace index for intermediate-peace countries that was in between that of the lower-peace and higher-peace, even though they were not in the training set. This study demonstrates how natural language processing and machine learning can help to generate new quantitative measures of social systems, which in this study, were linguistic differences resulting in a quantitative index of peace for countries at different levels of peacefulness.

## Introduction

### Importance of Language

Communication through language has been highlighted as the single most important process in constructing our reality (Luhmann, 1987; Karlberg, 2011). Language also plays a critical role in conflicts. The extreme power of “hate speech” to mobilize destruction and violence is evident around the globe. In Kenya, hate speech over social media and in blogs played a central role in inciting ethnic divides and conflict (Kimotho & Nyaga, 2016). In Nigeria, hate speech in the news was identified as a major driver of election violence (Ezeibe, 2021). Studies in Poland have shown that exposure to hate speech leads to lower evaluations of victims, greater distancing, and more outgroup prejudice (Soral, Bilewicz, & Winiewski, 2018). Peacekeepers working in conflict zones are currently using data science and natural language processing methods to track hate speech – monitoring hostile news accounts, blogs, and broadcast and social media posts in order to provide early warning predictions of increases in ethnic tensions or violence in local communities (Peace Tech Lab, 2020). These studies have focused on the prevention of destructive conflicts, approaching peace as the absence of harmful conflict.

However, highly peaceful societies have been found to evidence other conditions and processes in addition to an absence of violence that distinguish them from low peace nations, including the prevalence of non-warring norms, values and rituals (Fry et. al., 2021). Highly peaceful societies are also significantly more stable and have the lowest probability of lapsing into violence (Diehl, Goertz, & Gallegos, 2019). Yet, scarce research has been devoted to unpacking the conditions promoting higher levels of sustainable peace (Coleman & Deutsch, 2012; Coleman, et al., 2020). To build a foundation for sustainably peaceful societies, it is imperative to understand the drivers of peace. This has led to an increasing number of studies of “positive peace” (Fry, 2006; Deutsch & Coleman, 2016; Diehl, 2016; Goertz et al., 2016; Mahmoud & Makoond, 2017; Advanced Consortium of Cooperation, Conflict, and Complexity, 2018) to understand the active social forces that work together to generate and maintain peace in a society.

Linguistic features of peace and conflict can be found in all aspects of language including in phonology, grammar, semantics, pragmatics, and discourse (Bolivar, 2011). This leads us to ask, what are the properties of “peace speech” that are the other face of the coin from “hate speech”? Peace speech is a basic linguistic structure that may help to build and sustain peacefulness between people and between groups (Friedrich, 2007, 2019; Gomes de Matos, 2000; Ngabonziza, 2013). There is only limited empirical evidence identifying the specific features and effects of peace speech (Karlberg, 2011). As noted by peace linguist Patricia Friedrich (2019), “Just how much a change in vocabulary can shape the outcome of interactions should be a matter to be empirically verified by peace linguistics as soon as possible, so we can all move from the realm of possibility to the realm of empirical evidence and corroboration” (p. 120). Our aim here, is to use machine learning to identify some of the linguistic features of peace speech, namely the most frequently used words in lower-peace and higher-peace countries.

## Measuring Peace

There are several approaches to measuring peace. In some measures, peace is viewed as an objective state that can be defined, quantified, and measured according to a standardized set of parameters. This "technocratic" (Mac Ginty, 2013) approach assumes that peace consists of criteria that do not vary from case to case, and seeks to compare and rank cases, often in order to drive policy and funding (Fisher et al., 2020). Each of these indices consider a wide array of indicators that capture discrete elements of peacefulness and rank countries on their performance or attainment of these elements (Caplan, 2019). Prominent examples include the Global Peace Index (GPI) which measures peacefulness and its economic value (Institute for Economics and Peace, 2019a); the Positive Peace Index (PPI), which measures the conditions for peace in a society to flourish (Institute for Economics and Peace, 2019b; 2021); the Human Development Index (HDI), which measures a long and healthy life and a decent standard of living (United Nations Development Programme, 2021); the World Happiness Index (WHI), which measures happiness as perceived as by people themselves and their community (Helliwell, Layard & Sachs, 2019); and the Fragile States Index (FSI), which measures fragility, risk and vulnerability (Fund for Peace, 2019). The limitations of these indices include that they: a) are often based on incomplete data sets, b) are averaged across very different areas within nations to achieve national averages, c) based on vastly different assumptions and conceptualizations of what constitutes peacefulness, and d) are based on linear assumptions of cause and effect.

Alternative approaches hold that peace is highly context specific (Mac Ginty, 2013, Fisher et al., 2020). This local-centric approach centers on those who live within the context being measured, proposing that it is the members who live within a given society who should define what constitutes peace. This approach is more participatory, and seeks to include people from a given context to identify, define, and weight indicators of peace. Examples of this approach include the Everyday Peace Indicators (Firchow & Mac Ginty, 2017) and Generations for Peace (Yusuf & Voss, 2018). These methods help to address some of the limitations of top-down approaches to measuring peacefulness.

## Goals of this Study

Our primary goal here was to identify the words and their frequency of use in media articles that are most important in differentiating lower-peace and higher-peace countries. Certainly, words alone do not capture all the linguistic subtleties of language, but they can serve as a good starting point to explore the linguistic differences between lower-peace and higher-peace cultures. These words are the conduits of the social processes that underlie conflict and peace and may therefore provide insights into identifying those social processes.

Having developed a machine learning model to analyze media articles that accurately classified countries as lower-peace and higher-peace gave us the opportunity to use that model to develop a quantitative peace index, not only for the lower-peace and higher-peace countries in the training set, but also for other intermediate-peace countries that were not in the training set.

## Methods

### Overview

We used a data-driven approach. Over the previous centuries science has proceeded by using observations, experiments, data, and intuitions, to form theoretical frameworks that could then be supported or falsified by further experimental data. That is a top-down approach, from thoughts (theory) to data. In these studies, we went in the reverse direction, from the bottom-up, from data to thoughts (results) (A743, 2029; Reutter, 2020; Investopedia Team, 2022). Rather than conjecture what words differentiate conflicts from peace, and then search for the frequency of those words in the data, we used natural language processing and machine learning to analyze the words from lower-peace and higher-peace countries to find the set of words that best classifies a country as lower-peace or higher-peace.

An overview of the strategy used is shown in Figure 1. First, to focus on the differences between lower and higher peace countries, words likely to be common in both were removed by natural language processing. Also removed are names of people, places, and companies that would be confounding variables to predict the level of peace not related to language itself. Then the machine learning method is "trained" on countries of different levels of peace, that is, it is given some word data from those countries and then the parameters of the machine learning model are adjusted so that input into the model yields the correct output classification. Another set of "hyperparameters" on the how the machine learning algorithm works can also be adjusted, but those were held constant in the work presented here. This model is then "tested" by determining the statistical accuracy of its predictions given new word data from countries of different levels of peace. Next machine learning importance methods are used to find the words that are most important in the machine learning model in making its classification. This step determines the words that are most significant in differentiating lower and higher peace countries. This strategy to find the most important features in predicting the correct classification is typical of many applications of machine learning in data science and has been reliably used in other natural language applications, classifications based on features with numerical values, and image analysis (Lane, Howard, & Hapke, 2019; Raschka & Mirjalili, 2019). One machine learning model was trained on only lower-peace and higher-peace countries. It was then used to generate a quantitative machine learning peace index of any country, which could be in either the lower, intermediate, or higher-peace regimes. Therefore, it was applied to countries that are both similar and different to the countries that it was trained on.

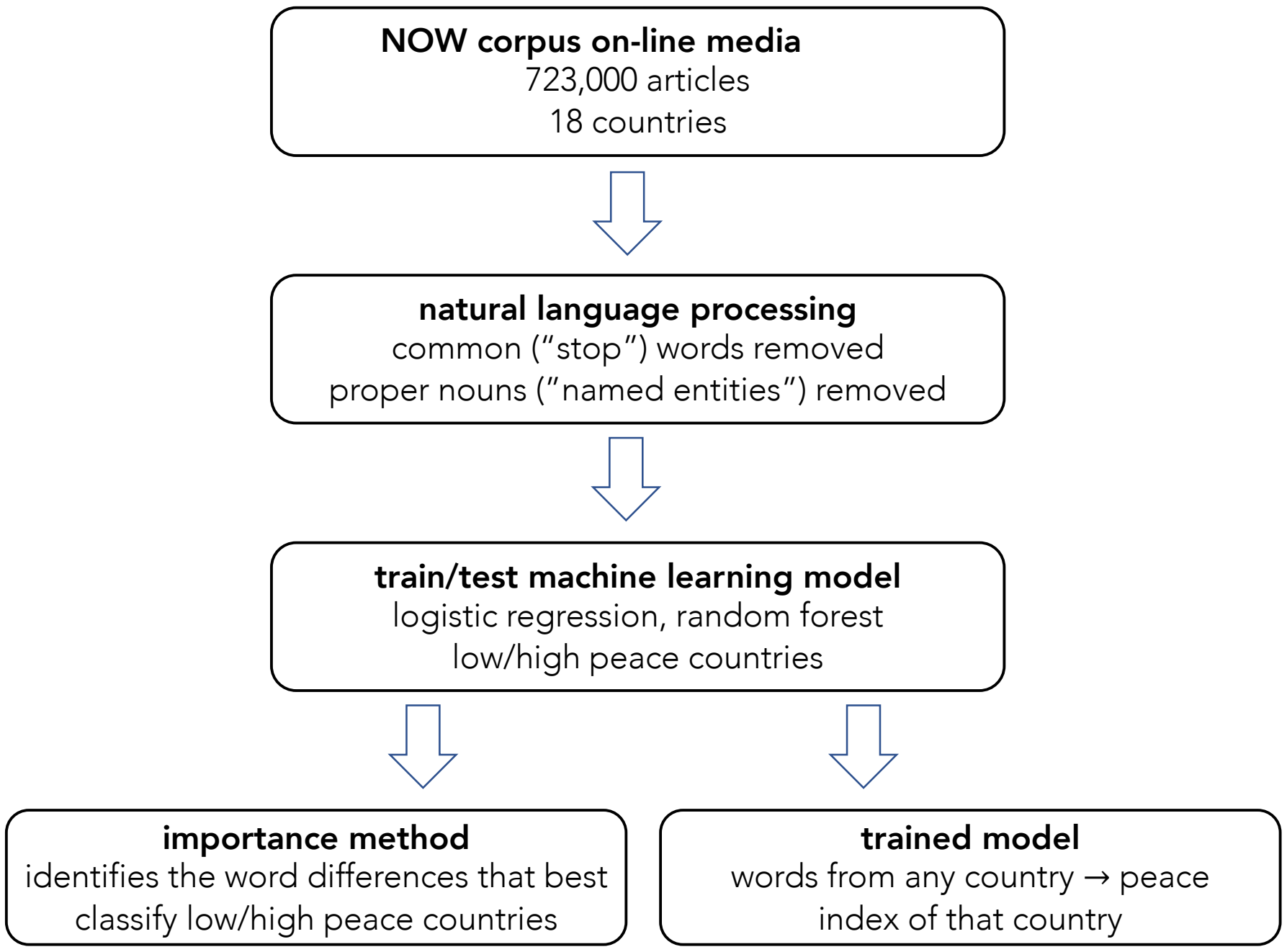

Figure 1. Strategy of the analysis.

## Data Collection and Pre-Processing

We used approximately 724,000 media articles published in English between January 2010 and September 2020, that had been collected from the News on the Web corpus (NOW, 2021). The NOW corpus includes articles from a very wide range of on-line newspapers, magazines, technical journals, and media broadcast stations. For example, a very small sample of articles from the United States includes the sources: AlterNet, Austin American-Statesman, Business Insider, Business Wire (press release), Chicago Tribune, FOX43.com, Jerusalem Post, Israel News, KCCI Des Moines, Kentwired, KOKI FOX 23, POWER magazine, Press of Atlantic City, The Jewish Press, USA TODAY, and Vulture. It was built by collecting data from such on-line sources. It is balanced in that in includes a wide variety of sources. It is biased in that all the sources are in English, which means it is best representative of countries where English is the native language, but less representative of countries where English is not the native language. Fairness is more difficult to ascertain, it is as fair, or as not fair, and the sources of its data. Since it has taken data from a wide variety of sources, it does include a wide variety of viewpoints.

**Table 1: Countries studied**

| | |
|---|---|
| 1 | Australia |
| 2 | Bangladesh |
| 3 | Canada |
| 4 | Ghana |
| 5 | Hong Kong |
| 6 | India |
| 7 | Ireland |
| 8 | Jamaica |
| 9 | Kenya |
| 10 | Malaysia |
| 11 | New Zealand |
| 12 | Nigeria |
| 13 | Philippines |
| 14 | Singapore |
| 15 | Sri Lanka |
| 16 | Tanzania |
| 17 | United Kingdom |
| 18 | United States |

Two countries from this corpus (Pakistan and South Africa) did not have a sufficient number of articles in English for further processing and were omitted from the machine learning model. All the stop words (common words), named entities (proper nouns such as names of people, places, and companies) and phrases unrelated to the article's content (such as ads) were removed by the methods of Jung et al. (2021) and manually as necessary. We then recorded the 300 most frequent words and their frequency of occurrence among all the articles combined for each country, which resulted in 767 unique words across all the 18 countries shown in Table 1.

## Machine Learning Models

Our strategy was to identify the words that are most important in a machine learning model in making the classification of the level of peace. To train a machine learning model required a training set of data with a known measure of peace in each country. However, as each existing peace index measures different social properties, there is no detailed agreement in the numerical values of these indices for each country. Therefore, there is no overall ordinal or ratio scale of the level of peace in each country to use in the training set. However, we can successfully use these indices, as described below, to group countries into three overall classes of lower-peace (class 0), higher-peace countries (class 1), and intermediate-peace (class 2), which we then used in our first 3-class machine learning model. As described in the results section below, that 3-class model was not very good at predicting the level of peace in a country. As there is more consensus amongst these indices for the countries that are at the extremes of lower-peace and higher-peace, we also developed a second, independent 2-class machine learning model using only the lower-peace (class 0) and higher-peace (class 1) countries. Using such extreme cases can help to clarify the differences between them. For example, Voukelatou et al. (2022) compared three peaceful countries (Portugal, Iceland, and New Zealand) with three of the most war-torn countries (DR Congo, Pakistan, and Yemen). This is called the Extreme Groups Approach (EGA) in psychology, where it must be used cautiously as to not artificially inflate statistical accuracy (Preacher, 2005). It is however appropriate and useful in standard machine learning to predict group membership, here to predict whether a country is lower-peace or higher-peace.

To determine the lower-peace and higher-peace countries, we first found the average values, over the years 2010-2019, for the GPI, PPI, WHI, FSI, and HDI indices for each country, as shown in Table 2. These indices were chosen as they are among the more prominent measures of levels of peace, conflict, and well-being at the country level. Each index also uses a different range from lower-peace to higher-peace to measure overall peace, respectively from 5-1, 5-1, 0-10, 120-0, and 0-1. To more easily compare them, as shown in Table 3, we linearly scaled the average of each index over the range 0-100, where 0 is lowest-peace and 100 is highest peace for the countries we analyzed. For each index, we then ordered each country by its average value and divided that list into thirds. The lower-peace countries were then defined as those with 3 or more of the 5 indices in the lowest group in that index, the higher-peace countries as those with 3 or more indices in the highest group in that index, and the intermediate-peace countries as those not in either group. We initially made this assignment for all 20 countries in the NOW corpus and kept that same assignment for the entire analysis, when later we removed two countries because they had a much smaller number of articles. Table 4 shows each of the five peace indices rates each country, the lower-peace group countries (in red): Bangladesh, Kenya, Nigeria, and Tanzania; the intermediate-peace group countries (in black): Ghana, Hong Kong, India, Jamaica, Malaysia, Philippines, Sri Lanka, and the United States; and the higher-peace group countries (in green): Australia, Canada, Ireland, New Zealand, Singapore, and the United Kingdom.

**Table 2: Peace indices from 5 sources. GPI=Global Peace Index, PPI=Positive Peace Index, WHI=World Happiness Index, FSI=Fragile States Index, and HDI=Human Development Index.**

| | Country | AVG GPI | AVG PPI | AVG WHI | AVG FSI | AVG HDI |
|---|---|---|---|---|---|---|
| 1 | Australia | 1.41 | 1.53 | 7.30 | 24.59 | 0.93 |
| 2 | Bangladesh | 2.11 | 3.62 | 4.66 | 91.76 | 0.59 |
| 3 | Canada | 1.37 | 1.50 | 7.40 | 24.93 | 0.91 |
| 4 | Ghana | 1.80 | 2.95 | 4.74 | 68.88 | 0.58 |
| 5 | Hong Kong | | | 5.48 | | 0.93 |
| 6 | India | 2.59 | 3.26 | 4.48 | 77.84 | 0.62 |
| 7 | Ireland | 1.43 | 1.38 | 7.01 | 23.58 | 0.93 |
| 8 | Jamaica | 2.12 | 2.53 | 5.55 | 64.99 | 0.72 |
| 9 | Kenya | 2.38 | 3.56 | 4.43 | 97.93 | 0.56 |
| 10 | Malaysia | 1.59 | 2.54 | 5.85 | 66.02 | 0.79 |
| 11 | New Zealand | 1.25 | 1.48 | 7.28 | 22.86 | 0.91 |
| 12 | Nigeria | 2.80 | 3.87 | 5.17 | 100.76 | 0.52 |
| 13 | Philippines | 2.50 | 3.28 | 5.24 | 84.73 | 0.70 |
| 14 | Singapore | 1.42 | 1.67 | 6.54 | 33.37 | 0.93 |
| 15 | Sri Lanka | 2.24 | 3.18 | 4.30 | 90.04 | 0.77 |
| 16 | Tanzania | 1.81 | 3.40 | 3.58 | 80.72 | 0.51 |
| 17 | United Kingdom | 1.84 | 1.55 | 6.90 | 34.08 | 0.92 |
| 18 | United States | 2.29 | 1.79 | 7.03 | 35.45 | 0.92 |

**Table 3: Peace indices in Table 2 each independently, linearly scaled for these countries where 0 is lowest-peace and 100 is highest peace.**

| | Country | AVG GPI | AVG PPI | AVG WHI | AVG FSI | AVG HDI |
|---|---|---|---|---|---|---|
| 1 | Australia | 89.68 | 93.98 | 97.38 | 97.78 | 100.00 |
| 2 | Bangladesh | 44.52 | 10.04 | 28.27 | 11.55 | 19.05 |
| 3 | Canada | 92.26 | 95.18 | 100.00 | 97.34 | 95.24 |
| 4 | Ghana | 64.52 | 36.95 | 30.37 | 40.92 | 16.67 |
| 5 | Hong Kong | | | 49.74 | | 100.00 |
| 6 | India | 13.55 | 24.50 | 23.56 | 29.42 | 26.19 |
| 7 | Ireland | 88.39 | 100.00 | 89.79 | 99.08 | 100.00 |
| 8 | Jamaica | 43.87 | 53.82 | 51.57 | 45.92 | 50.00 |
| 9 | Kenya | 27.10 | 12.45 | 22.25 | 3.63 | 11.90 |
| 10 | Malaysia | 78.06 | 53.41 | 59.42 | 44.60 | 66.67 |
| 11 | New Zealand | 100.00 | 95.98 | 96.86 | 100.00 | 95.24 |
| 12 | Nigeria | 0.00 | 0.00 | 41.62 | 0.00 | 2.38 |
| 13 | Philippines | 19.35 | 23.69 | 43.46 | 20.58 | 45.24 |
| 14 | Singapore | 89.03 | 88.35 | 77.49 | 86.51 | 100.00 |
| 15 | Sri Lanka | 36.13 | 27.71 | 18.85 | 13.76 | 61.90 |
| 16 | Tanzania | 63.87 | 18.88 | 0.00 | 25.73 | 0.00 |
| 17 | United Kingdom | 61.94 | 93.17 | 86.91 | 85.60 | 97.62 |
| 18 | United States | 32.90 | 83.53 | 90.31 | 83.84 | 97.62 |

**Table 4: Table 3 color coded independently for each index, lower-peace group=red, higher-peace group=green, and intermediate-peace group=yellow. Countries were identified as lower-peace if they had 3 or more peace indices in the lowest group, higher-peace if they had 3 or more indices in the highest group, and the remaining countries as intermediate-peace.**

| | Country | AVG GPI | AVG PPI | AVG WHI | AVG FSI | AVG HDI |
|---|---|---|---|---|---|---|
| 1 | **Australia** | 89.68 | 93.98 | 97.38 | 97.78 | 100.00 |
| 2 | **Bangladesh** | 44.52 | 10.04 | 28.27 | 11.55 | 19.05 |
| 3 | **Canada** | 92.26 | 95.18 | 100.00 | 97.34 | 95.24 |
| 4 | Ghana | 64.52 | 36.95 | 30.37 | 40.92 | 16.67 |
| 5 | Hong Kong | | | 49.74 | | 100.00 |
| 6 | India | 13.55 | 24.50 | 23.56 | 29.42 | 26.19 |
| 7 | **Ireland** | 88.39 | 100.00 | 89.79 | 99.08 | 100.00 |
| 8 | Jamaica | 43.87 | 53.82 | 51.57 | 45.92 | 50.00 |
| 9 | **Kenya** | 27.10 | 12.45 | 22.25 | 3.63 | 11.90 |
| 10 | Malaysia | 78.06 | 53.41 | 59.42 | 44.60 | 66.67 |
| 11 | **New Zealand** | 100.00 | 95.98 | 96.86 | 100.00 | 95.24 |
| 12 | **Nigeria** | 0.00 | 0.00 | 41.62 | 0.00 | 2.38 |
| 13 | Philippines | 19.35 | 23.69 | 43.46 | 20.58 | 45.24 |
| 14 | **Singapore** | 89.03 | 88.35 | 77.49 | 86.51 | 100.00 |
| 15 | Sri Lanka | 36.13 | 27.71 | 18.85 | 13.76 | 61.90 |
| 16 | **Tanzania** | 63.87 | 18.88 | 0.00 | 25.73 | 0.00 |
| 17 | **United Kingdom** | 61.94 | 93.17 | 86.91 | 85.60 | 97.62 |
| 18 | United States | 32.90 | 83.53 | 90.31 | 83.84 | 97.62 |

Table 5 shows the number of articles and words in the data from the countries used for the 3-class model of the lower-peace, intermediate-peace, and higher-peace countries. Table 6 shows that data for the 2-class model of the lower-peace and higher-peace countries.

**Table 5: Data for the countries of the 3-class training set, lower-peace=red, intermediate-peace=black, and higher-peace=green.**

| | Country | Number of Articles | Number of Words | Per Cent Articles | Per cent Words |
|---|---|---|---|---|---|
| 1 | **Bangladesh** | **15,245** | **1,183,478** | 2.11 | 2.05 |
| 2 | **Kenya** | **30,694** | **1,940,412** | 4.24 | 3.36 |
| 3 | **Nigeria** | **52,895** | **5,307,297** | 7.31 | 9.18 |
| 4 | **Tanzania** | **6,164** | **500,702** | 0.85 | 0.87 |
| | total | 104,998 | 8,931,889 | 14.51 | 15.45 |
| | | | | | |
| 1 | Ghana | 22,783 | 1,699,258 | 3.15 | 2.94 |
| 2 | Hong Kong | 2,301 | 253,881 | 0.32 | 0.44 |
| 3 | India | 76,555 | 5,294,277 | 10.58 | 9.16 |
| 4 | Jamaica | 33,401 | 2,580,436 | 4.62 | 4.46 |
| 5 | Malaysia | 30,394 | 2,210,748 | 4.20 | 3.82 |
| 6 | Philippines | 61,474 | 4,166,281 | 8.50 | 7.21 |
| 7 | Sri Lanka | 11,329 | 983,349 | 1.57 | 1.70 |
| 8 | United States | 67,406 | 5,975,558 | 9.32 | 10.33 |
| | total | 305,643 | 23,163,788 | 42.24 | 40.06 |
| | | | | | |
| 1 | **Australia** | **62,683** | **5,599,285** | 8.66 | 9.68 |
| 2 | **Canada** | **73,869** | **6,981,358** | 10.21 | 12.07 |
| 3 | **Ireland** | **60,190** | **4,293,895** | 8.32 | 7.43 |
| 4 | **New Zealand** | **56,483** | **4,417,416** | 7.81 | 7.64 |
| 5 | **Singapore** | **20,195** | **1,345,811** | 2.79 | 2.33 |
| 6 | **United Kingdom** | **39,513** | **3,085,992** | 5.46 | 5.34 |
| | total | 312,933 | 25,723,757 | 43.25 | 44.49 |

We used the random forest and logistic regression classifiers (Pedregosa et al., 2011a,b) to train and test the 3-class model of lower-peace (class 0), higher-peace (class 1), and intermediate-peace (class 2) and independently the 2-class model of lower-peace (class 0) and higher-peace (class 1). In all cases the features were the 767 most frequently used words across all the countries in the data. There are different ways to both train and test such models (Raschka & Mirjalili, 2019). As typically done, we first trained each model by using 80% of the data and then tested it on the remaining 20% of the data. We also used a cross-validation method (cross-validation, 2023) where the model is trained on all but one country, tested on the excluded country, and this is repeated for each different country being excluded. This makes more efficient use of the information in the data but requires additional computational time for the repeated trainings.

**Table 6: Data for the countries of the 2-class training set, lower-peace=red and higher-peace=green.**

| | Country | Number of Articles | Number of Words | Per Cent Articles | Per cent Words |
|---|---|---|---|---|---|
| 1 | **Bangladesh** | **15,245** | **1,183,478** | **3.65** | **3.41** |
| 2 | **Kenya** | **30,694** | **1,940,412** | **7.34** | **5.60** |
| **3** | **Nigeria** | **52,895** | **5,307,297** | **12.66** | **15.31** |
| **4** | **Tanzania** | **6,164** | **500,702** | **1.47** | **1.44** |
| | total | 104,998 | 8,931,889 | 25.12 | 25.77 |
| | | | | | |
| 1 | **Australia** | **62,683** | **5,599,285** | **15.00** | **16.16** |
| 2 | **Canada** | **73,869** | **6,981,358** | **17.67** | **20.14** |
| 3 | **Ireland** | **60,190** | **4,293,895** | **14.40** | **12.39** |
| 4 | **New Zealand** | **56,483** | **4,417,416** | **13.51** | **12.75** |
| 5 | **Singapore** | **20,195** | **1,345,811** | **4.83** | **3.88** |
| 6 | **United Kingdom** | **39,513** | **3,085,992** | **9.45** | **8.90** |
| | total | 312,933 | 25,723,757 | 74.88 | 74.23 |

## Machine Learning Peace Index

As shown in Figure 1, the top-down machine learning approach leads from the data through natural language processing, to the training and testing of the machine learning model. We then used the machine learning model in two different ways. First, as already described, we used its importance methods to identify the word differences that best classify (that is, predict) which countries are lower or higher peace. Second, we also used it to provide a quantitative peace index from the media data.

A regression model (such as logistic regression but not random forest), trained on only a set of classes, can determine the parameters of mathematical equations that best predict the probability of membership in each class. From the 2-class model, trained on only the word frequencies in the lower-peace and higher-peace countries, these equations determined the probability, p, that a country is in the higher-peace class (class 1). In the binary classification task, the model classifies countries as lower peace if $p < 0.5$ and higher peace if $p \geq 0.5$. We used this value of p, which is a quantitative measure of the probability of being in the higher-peace class (class 1), as a measure of the level of peace in any country in either the lower, intermediate, or higher-peace regimes. Therefore, it was applied to countries that are both similar and different to the countries that it was trained on. To be consistent with our scaling of the other peace indices in Table 3, we defined a machine learning peace index as 100 x p, so that 0 is lowest peace and 100 is highest peace.

## Results

### Performance Measures

Table 7 shows the mean ± sem of the performance measures for random guessing and the random forest and logistic regression classifiers on the 3-class and 2-class models. These are the Accuracy=(TP+TN)/(FP+FN+TP+TN), the Precision=TP/(TP+FP), the Recall=TP/(FN+TP), and F1=2(Precision x Recall)/(Precision+Recall), where TP=true positive, TN=true negative, FP=false positive, and FN=false negative.

**Table 7: Performance measures over 20 runs of each machine learning method, each value is mean ± sem. 3-class models are: lower-peace, intermediate-peace, and higher-peace. 2-class models are: lower-peace and higher-peace.**

| | **Accuracy** | **Precision** | **Recall** | **F1** |
|---|---|---|---|---|
| 3-class, Random Guessing | 0.356 ± 0.033 | 0.368 ± 0.034 | 0.356 ± 0.033 | 0.354 ± 0.033 |
| **Random Forest** | | | | |
| 3-class, 80/20 train/test | 0.525 ± 0.040 | 0.420 ± 0.056 | 0.525 ± 0.040 | 0.437 ± 0.048 |
| 3-class, 17 to predict one | 0.567 ± 0.015 | 0.500 ± 0.018 | 0.567 ± 0.015 | 0.525 ± 0.015 |
| 2-class, 9 to predict one | 0.960 ± 0.013 | 0.965 ± 0.012 | 0.960 ± 0.013 | 0.960 ± 0.013 |
| **Logistic Regression** | | | | |
| 3-class, 80/20 train/test | 0.388 ± 0.053 | 0.238 ± 0.057 | 0.388 ± 0.053 | 0.272 ± 0.050 |
| 3-class, 17 to predict one | 0.611 * | 0.520 * | 0.611 * | 0.558 * |
| 2-class, 9 to predict one | 1.000 * | 1.000 * | 1.000 * | 1.000 * |

*Since the logistic regression converges to the same values on each run, sem = 0 for these values.

First we consider the results from the 3-class model using the lower-peace, intermediate-peace, and higher-peace countries. Random guessing of the three classes, averaged over 20 guesses, yielded an accuracy of 0.356, within the error expected around 0.333. The 80/20 train/test split using all 18 countries, was only weakly more successful, with accuracies of 0.525 for the random forest and 0.388 for logistic regression. Nonetheless, for example, for the random forest, this accuracy is still statistically significantly greater than random guessing ($Z = (\Delta\text{means})/\text{sem} = 4.14$, $p < 2.0 \times 10^{-5}$, one-tailed).

We had expected that the more efficient cross validation using all the data from 17 countries to predict the class of the one country not used in each training, would significantly improve the accuracy. This was not the case. For random forest the accuracy only improved slightly from 0.525 to 0.567 and for logistic regression the accuracy increased a little more from 0.388 to 0.611.

Second, we now consider the results from the 2-class model using only the lower-peace and higher-peace countries. Now the cross-validation model dramatically increased the accuracy to 0.960 for the random forest model and 1.000 for the logistic regression model. This dramatic improvement in the accuracy of the 2-class model over the 3-class model makes sense in the following way. Since the peace indices of the humans disagree with each other how can our machine learning model figure it out? But, the peace indices of the humans are more aligned with each other for the countries that are extremely low or high peace, so they are pretty sure which are the most lower-peace and higher-peace countries. Thus, training on the 2-class model using only the lower-peace and higher-peace countries makes it possible for the machine learning model to properly associate the level of peace with the word frequencies.

The number of articles and words from each country is shown in Table 6. The excellent values for the performance measures in Table 7 for the 2-class model demonstrate that the differences in numbers of articles or words between the countries had no significant negative impact on these results. For example, the 20 random forest runs for the 2-class model there were no mis-classifications for 188 classifications, only 5 times were higher-peace countries mis-classified as lower-peace, and only 7 times were lower-peace countries mis-classified as higher-peace. The 20 logistic regression runs all converged to the same values with no mis-classifications.

NOTE: Because the performance measures of the 2-class model were so much better than that of the 3-class model, all the following results are based on the 2-class model.

## Most Frequent and Important Words in Lower-Peace and Higher-Peace Countries

The machine learning model was used to find the words most important in differentiating lower-peace and higher-peace countries. Figure 2 shows the words that were most frequently used in the articles in the lower-peace and higher-peace countries. We also used the *feature_importances* method from the random forest classifier to determine which of these words were most important in correctly predicting whether a country is lower-peace or higher-peace. The highest frequency words were more likely the words of the highest feature importance, but interestingly, many words of lower frequency were also important in predicting whether a country was lower-peace or higher-peace. Figure 2 shows the 100 most frequently used words in higher-peace and lower-peace countries with the words of highest feature importance highlighted in yellow.

Figure 2. The 100 most frequent words for the higher-peace and lower-peace countries. Yellow indicates the words of highest feature importance in making the higher-peace and lower-peace classification by the random forest feature importance method.

| | High Peace | Count | | High Peace | Count | | | Low Peace | Count | | Low Peace | Count |
|---|---|---|---|---|---|---|---|---|---|---|---|---|
| 1 | **time** | 4.10E+05 | 51 | lead | 1.16E+05 | | 1 | **state** | 1.88E+05 | 51 | power | 4.07E+04 |
| 2 | people | 3.78E+05 | 52 | public | 1.15E+05 | | 2 | **governme** | 1.61E+05 | 52 | **security** | 4.05E+04 |
| 3 | new | 3.13E+05 | 53 | number | 1.15E+05 | | 3 | people | 1.37E+05 | 53 | group | 4.03E+04 |
| 4 | **work** | 3.09E+05 | 54 | **child** | 1.14E+05 | | 4 | **country** | 1.35E+05 | 54 | support | 4.01E+04 |
| 5 | use | 3.04E+05 | 55 | school | 1.13E+05 | | 5 | **president** | 9.33E+04 | 55 | federal | 3.96E+04 |
| 6 | **like** | 2.91E+05 | 56 | set | 1.13E+05 | | 6 | **time** | 9.12E+04 | 56 | day | 3.94E+04 |
| 7 | come | 2.68E+05 | 57 | woman | 1.09E+05 | | 7 | come | 8.10E+04 | 57 | **start** | 3.92E+04 |
| 8 | need | 2.09E+05 | 58 | share | 1.08E+05 | | 8 | **work** | 6.93E+04 | 58 | local | 3.92E+04 |
| 9 | **look** | 2.04E+05 | 59 | **run** | 1.08E+05 | | 9 | new | 6.85E+04 | 59 | **place** | 3.86E+04 |
| 10 | include | 2.01E+05 | 60 | issue | 1.06E+05 | | 10 | use | 6.76E+04 | 60 | **sector** | 3.78E+04 |
| 11 | know | 1.93E+05 | 61 | **try** | 1.06E+05 | | 11 | **national** | 6.73E+04 | 61 | money | 3.70E+04 |
| 12 | **want** | 1.88E+05 | 62 | lot | 1.05E+05 | | 12 | need | 6.41E+04 | 62 | **provide** | 3.70E+04 |
| 13 | **way** | 1.86E+05 | 63 | told | 1.04E+05 | | 13 | **like** | 6.28E+04 | 63 | help | 3.57E+04 |
| 14 | company | 1.84E+05 | 64 | case | 1.02E+05 | | 14 | know | 6.03E+04 | 64 | nation | 3.54E+04 |
| 15 | **governme** | 1.82E+05 | 65 | comment | 1.00E+05 | | 15 | police | 5.60E+04 | 65 | office | 3.51E+04 |
| 16 | **game** | 1.73E+05 | 66 | **state** | 9.94E+04 | | 16 | service | 5.37E+04 | 66 | chief | 3.48E+04 |
| 17 | **think** | 1.71E+05 | 67 | area | 9.80E+04 | | 17 | include | 5.35E+04 | 67 | ensure | 3.48E+04 |
| 18 | **good** | 1.68E+05 | 68 | player | 9.80E+04 | | 18 | high | 5.32E+04 | 68 | **internatio** | 3.42E+04 |
| 19 | **world** | 1.68E+05 | 69 | local | 9.79E+04 | | 19 | **party** | 5.31E+04 | 69 | number | 3.41E+04 |
| 20 | **team** | 1.65E+05 | 70 | **found** | 9.71E+04 | | 20 | governor | 5.31E+04 | 70 | road | 3.40E+04 |
| 21 | high | 1.64E+05 | 71 | **add** | 9.67E+04 | | 21 | public | 5.31E+04 | 71 | **act** | 3.35E+04 |
| 22 | **home** | 1.60E+05 | 72 | health | 9.61E+04 | | 22 | issue | 5.23E+04 | 72 | told | 3.33E+04 |
| 23 | right | 1.55E+05 | 73 | base | 9.58E+04 | | 23 | company | 5.19E+04 | 73 | set | 3.32E+04 |
| 24 | change | 1.52E+05 | 74 | **site** | 9.49E+04 | | 24 | **world** | 5.14E+04 | 74 | health | 3.32E+04 |
| 25 | help | 1.52E+05 | 75 | follow | 9.48E+04 | | 25 | **member** | 5.14E+04 | 75 | communit | 3.31E+04 |
| 26 | business | 1.50E+05 | 76 | police | 9.47E+04 | | 26 | **election** | 5.02E+04 | 76 | **thing** | 3.30E+04 |
| 27 | life | 1.48E+05 | 77 | plan | 9.26E+04 | | 27 | developme | 5.01E+04 | 77 | **order** | 3.27E+04 |
| 28 | day | 1.47E+05 | 78 | win | 9.19E+04 | | 28 | report | 4.96E+04 | 78 | increase | 3.17E+04 |
| 29 | **start** | 1.44E+05 | 79 | ask | 9.02E+04 | | 29 | **court** | 4.96E+04 | 79 | **home** | 3.17E+04 |
| 30 | service | 1.41E+05 | 80 | **find** | 8.88E+04 | | 30 | **want** | 4.89E+04 | 80 | **team** | 3.17E+04 |
| 31 | **thing** | 1.41E+05 | 81 | **story** | 8.86E+04 | | 31 | business | 4.86E+04 | 81 | **look** | 3.16E+04 |
| 32 | family | 1.38E+05 | 82 | cost | 8.85E+04 | | 32 | **bank** | 4.74E+04 | 82 | house | 3.16E+04 |
| 33 | **place** | 1.37E+05 | 83 | increase | 8.84E+04 | | 33 | **way** | 4.74E+04 | 83 | **director** | 3.15E+04 |
| 34 | **country** | 1.36E+05 | 84 | **member** | 8.80E+04 | | 34 | **good** | 4.73E+04 | 84 | **according** | 3.14E+04 |
| 35 | **play** | 1.35E+05 | 85 | man | 8.79E+04 | | 35 | area | 4.70E+04 | 85 | education | 3.12E+04 |
| 36 | **big** | 1.34E+05 | 86 | event | 8.71E+04 | | 36 | life | 4.63E+04 | 86 | follow | 3.10E+04 |
| 37 | report | 1.33E+05 | 87 | young | 8.70E+04 | | 37 | school | 4.59E+04 | 87 | **officer** | 3.06E+04 |
| 38 | market | 1.31E+05 | 88 | news | 8.61E+04 | | 38 | case | 4.45E+04 | 88 | end | 3.06E+04 |
| 39 | **informatio** | 1.26E+05 | 89 | **national** | 8.53E+04 | | 39 | **political** | 4.38E+04 | 89 | system | 3.04E+04 |
| 40 | **great** | 1.26E+05 | 90 | **open** | 8.52E+04 | | 40 | lead | 4.37E+04 | 90 | **economic** | 3.03E+04 |
| 41 | **provide** | 1.25E+05 | 91 | minister | 8.49E+04 | | 41 | **general** | 4.35E+04 | 91 | **university** | 3.02E+04 |
| 42 | point | 1.24E+05 | 92 | system | 8.45E+04 | | 42 | market | 4.28E+04 | 92 | change | 3.01E+04 |
| 43 | support | 1.22E+05 | 93 | price | 8.42E+04 | | 43 | **project** | 4.27E+04 | 93 | family | 2.98E+04 |
| 44 | **year** | 1.22E+05 | 94 | result | 8.42E+04 | | 44 | minister | 4.26E+04 | 94 | man | 2.96E+04 |
| 45 | city | 1.22E+05 | 95 | level | 8.39E+04 | | 45 | **add** | 4.21E+04 | 95 | point | 2.95E+04 |
| 46 | best | 1.21E+05 | 96 | **believe** | 8.34E+04 | | 46 | **law** | 4.20E+04 | 96 | industry | 2.92E+04 |
| 47 | long | 1.21E+05 | 97 | continue | 8.31E+04 | | 47 | right | 4.16E+04 | 97 | student | 2.90E+04 |
| 48 | communit | 1.20E+05 | 98 | **experience** | 8.21E+04 | | 48 | woman | 4.16E+04 | 98 | **chairman** | 2.88E+04 |
| 49 | end | 1.20E+05 | 99 | expect | 8.17E+04 | | 49 | **child** | 4.10E+04 | 99 | oil | 2.88E+04 |
| 50 | group | 1.17E+05 | 100 | term | 8.15E+04 | | 50 | leader | 4.08E+04 | 100 | fund | 2.85E+04 |

A decision tree graph is a good method to classify data, but it can learn to fit the training data so tightly that it does not generalize to properly classify new data. The random forest classifier was developed to avoid such overfitting by creating many different classification trees from random subsets of the data, literally a random forest, to make the classification. Hence, each time it is run, the model will identify a slightly different set of words that best classifies a country as lower-peace and higher-peace. We found that the words of high frequency and high importance were very similar in each run, but that there was more variation in the words identified of lower-frequency and lower-importance.

Figures 3 and 4 show word clouds of the words of highest feature importance, with the size of those words scaled to their frequency of occurrence, in green for higher-peace countries and in red for lower-peace countries. The word cloud in Figure 3 for the higher-peace countries is dominated by words associated with positive sentiments and play: “time” “like”, “game”, “play”, “good”, “team”. On the other hand, Figure 4 for the lower-peace countries, is dominated by words about social structures, such as: “state”, “government”, “country” and governmental control such as: “court”, “general”, “law”. A preliminary and speculative analysis of these results suggests that lower-peace countries are characterized by words of government control and fear. The direction of the arrow of causality is not clear. Do the social realities lead to these words, or do these words lead to the social realities? Can “peace speech” in news and social media enhance the prospects for peace or only reflect it?

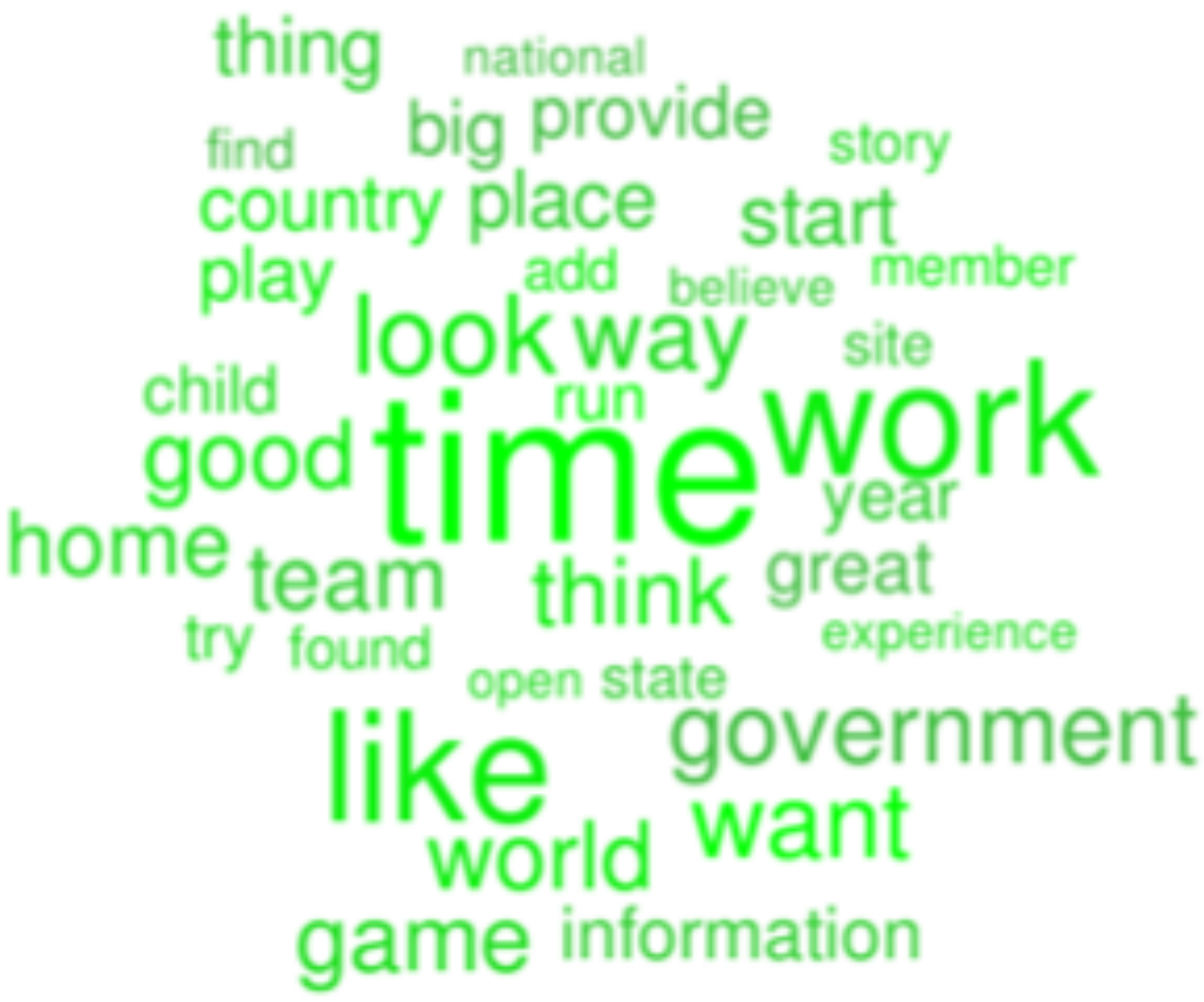

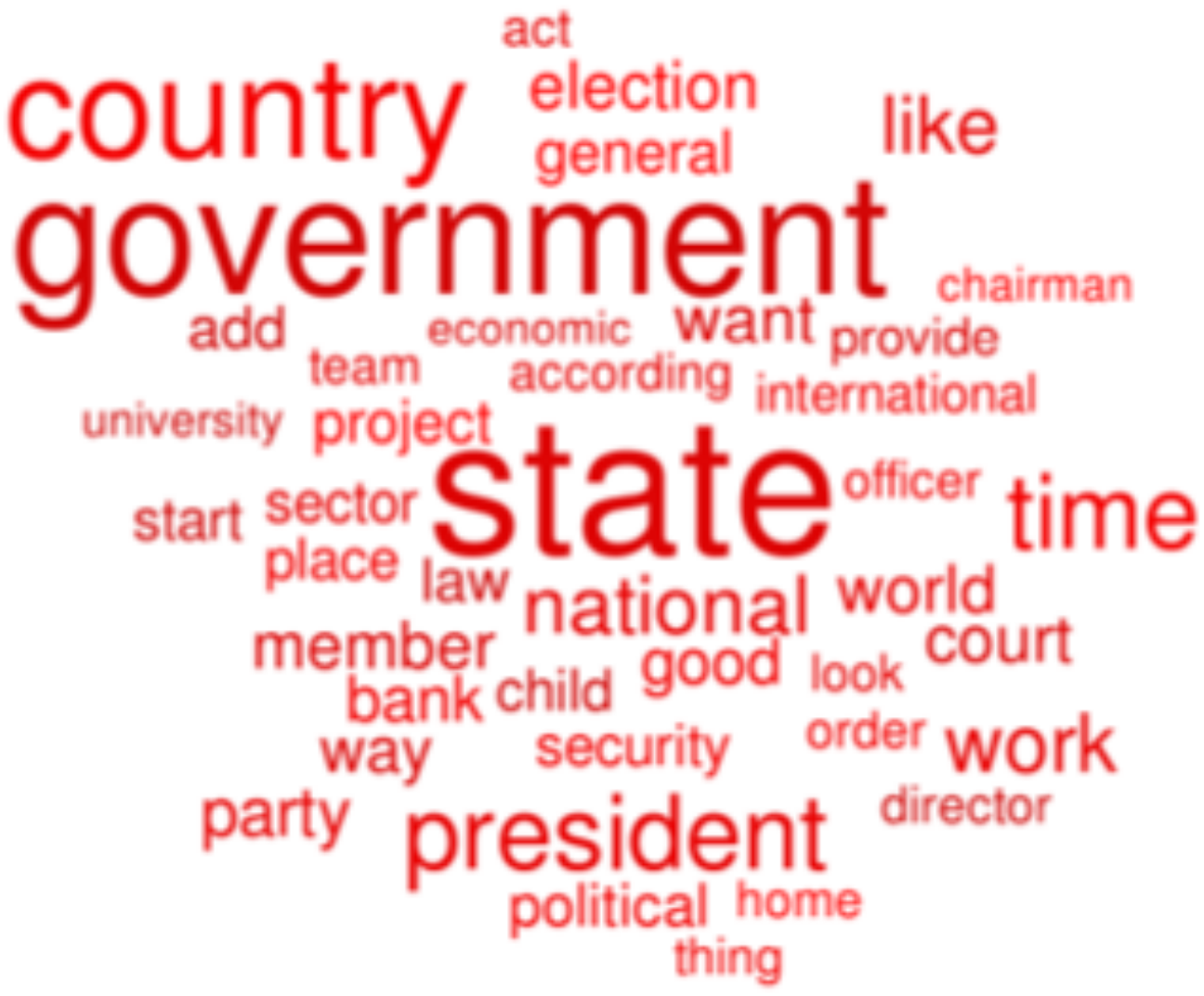

Figure 3. Green word cloud of the words of highest feature importance, with their size scaled to their frequency of occurrence, for higher-peace countries.

Figure 4. Red word cloud of the words of highest feature importance, with their size scaled to their frequency of occurrence, for lower-peace countries

## Machine Learning Peace Index

A machine learning model trained by the logistic regression classifier using word frequencies in media articles to classify countries as lower-peace and higher-peace, was also used to determine a quantitative peace index, not only for the lower-peace and higher-peace countries in the training set, but also for other intermediate-peace countries that were not in the training set. The logistic regression model used the word frequencies from each country to compute the probability p that country was higher-peace (class 1). The machine learning peace index was then equal to 100 times p. Table 8 shows the 18 countries in rank order of this machine learning peace index compared to the GPI, PPI, WHI, FSI, and HDI indices. There are two important findings from this computation.

**Table 8: Machine learning (ML) peace index compared to the other peace indices. Training set of countries: lower-peace=red and higher-peace=green.**

| | Country | ML Peace Index | AVG GPI | AVG PPI | AVG WHI | AVG FSI | AVG HDI |
|---|---|---|---|---|---|---|---|
| 1 | **Tanzania** | **6.22** | 63.87 | 18.88 | 0.00 | 25.73 | 0.00 |
| 2 | **Nigeria** | **6.30** | 0.00 | 0.00 | 41.62 | 0.00 | 2.38 |
| 3 | **Bangladesh** | **9.56** | 44.52 | 10.04 | 28.27 | 11.55 | 19.05 |
| 4 | Sri Lanka | **12.69** | 36.13 | 27.71 | 18.85 | 13.76 | 61.90 |
| 5 | Ghana | **13.61** | 64.52 | 36.95 | 30.37 | 40.92 | 16.67 |
| 6 | **Kenya** | **14.31** | 27.10 | 12.45 | 22.25 | 3.63 | 11.90 |
| 7 | Jamaica | **43.71** | 43.87 | 53.82 | 51.57 | 45.92 | 50.00 |
| 8 | Malaysia | **49.42** | 78.06 | 53.41 | 59.42 | 44.60 | 66.67 |
| 9 | Philippines | **53.78** | 19.35 | 23.69 | 43.46 | 20.58 | 45.24 |
| 10 | India | **56.45** | 13.55 | 24.50 | 23.56 | 29.42 | 26.19 |
| 11 | Hong Kong | **57.99** | | | 49.74 | | 100.00 |
| 12 | **Singapore** | **90.38** | 89.03 | 88.35 | 77.49 | 86.51 | 100.00 |
| 13 | **New Zealand** | **92.50** | 100.00 | 95.98 | 96.86 | 100.00 | 95.24 |
| 14 | United States | **94.01** | 32.90 | 83.53 | 90.31 | 83.84 | 97.62 |
| 15 | **Canada** | **94.47** | 92.26 | 95.18 | 100.00 | 97.34 | 95.24 |
| 16 | **United Kingdom** | **94.47** | 61.94 | 93.17 | 86.91 | 85.60 | 97.62 |
| 17 | **Ireland** | **95.87** | 88.39 | 100.00 | 89.79 | 99.08 | 100.00 |
| 18 | **Australia** | **95.91** | 89.68 | 93.98 | 97.38 | 97.78 | 100.00 |

First, as can be seen in Table 8, although the model was trained only on the lower-peace and higher-peace countries, and it has never been given any data whatsoever about the intermediate-peace countries, it correctly ranks those intermediate-peace countries in between the lowest lower-peace and highest higher-peace countries. This striking result confirms that the machine learning model has learned something real and substantive from the word frequencies in the lower-peace and higher-peace countries, that correctly generalizes to the intermediate-peace countries.

Second, unlike the positivist approaches to measuring peace, which a priori choose social indicators from their conceptualization of peace, the machine learning peace index is data driven and free of any assumptions about which words or their frequencies are most representative of peace. The choices of which words, and their frequencies, are important in measuring peace, arise solely from training the machine model, with samples of media articles from countries identified as lower-peace and higher-peace. This is a new and valuable data driven, bottom-up approach. It is the reverse of a classical top-down approach where a conceptual framework is used to hypothesize which words best measure peace and then test that hypothesis with data. As previous measures of peace had used a top-down approach based on a priori assumptions, here we explored what could additional be learned by using a bottom-up, data driven, data science approach. We chose to explore that bottom-up approach because it could, and in fact did here, provide us new insights into the differences in language between lower and higher peace countries that had never before been formulated into hypotheses to be tested. Every flower of a different color adds beauty to the garden.

## Discussion

The language that we use to communicate across our differences both reflects our internal view of the world and influences our external world. "Hate speech" can mobilize violence and destruction. Much less is known about "peace speech" that characterizes peaceful cultures and that may also help to generate or sustain peace. Our long-range aim is to identify the linguistic features of speech that characterizes lower-peace and higher-peace societies. In this study, we identified the words in media articles most associated with lower-peace and higher-peace countries. Certainly, words alone do not capture all the linguistic subtleties of language, but they can serve as a good starting point to explore the linguistic differences between lower-peace and higher-peace cultures.

We used a novel data science approach to identify those words. These data science methods, developed in computer science, which are widely used in commerce, are now being increasingly applied to gain new understanding of systems in the physical, biological, medical, and social sciences. A classical approach would be to use theoretical concepts to generate sets of words expected to be more frequently found in lower-peace and higher-peace countries and then test whether that is indeed the case. Instead, we used a modern data science approach to identify those words that are the most important in predicting whether a country is lower-peace or higher-peace.

From the machine learning model, we found that the words that are most important in differentiating lower-peace and higher-peace countries are those shown in Figures 3 and 4. These words suggest that lower-peace countries are characterized by words of government, order, control and fear (e.g. government, state, court), while higher-peace countries are characterized by words of optimism for the future and fun (e.g. time, like, game). Words are both a cause and a consequence of the social processes that lead to lower or higher levels of peace. The link between these words and their associated social processes needs to be developed further. Having identified those words, at least, provides a starting point for that exploration.

Having trained the machine learning model to use words to recognize the differences between lower-peace and higher-peace countries gave us the opportunity to rank countries on their level of peace, as shown in Table 8. Current peace indices use conceptual frameworks to choose data believed to be indicators of peace. Our machine learning peace index is agnostic to such theoretical assumptions or frameworks. The parameters of the machine learning model arise only from its ability to use word frequencies to correctly classify countries as lower-peace or higher-peace. How does our quantitative machine learning peace index compare to other measures of peace? As can be seen from Table 8, the overall ranking of countries by our machine learning peace index is similar to the overall rankings by five other peace indices based on their theoretical conceptual frameworks. This comparison cannot be exact, as each of those other indices differs somewhat from each other. Recently, other machine learning methods have been used to show that events from the GDELT (Global Data on Events, Locations, and Tone ) digital news database (Leetaru, 2013) successfully correlates with, and can even predict, the values of the GPI over time (Voukelaton et al. 2020; 2022). Those studies used pre-assigned event categories, while our work here used machine learning to identify the words that differentiate lower and higher levels of peace without prior assumptions on what those words would be. Our approach, for example, has led to the unanticipated finding that news stories about “games”, “teams”, and “play” are representative of higher levels of peace.

In order to avoid difficulties in translation, we restricted our analysis to sources in English. This means that the data we analyzed may reflect a Western bias in the countries chosen because those countries have the most extensive news media in English, and as many of the higher peace countries are in the Global North, while the lower peace countries are in the Global South. That may influence the words determined from the lower-peace and higher-peace countries and the quantitative values of the machine learning model peace index.

Future directions for these studies include identifying the social processes reflected in the different sets of words in the lower-peace and higher-peace countries. One promising approach is to use the word frequencies to identify societies with “‘tight’ cultural groups that have strong norms and little tolerance for deviance while other ‘loose’ groups that have weaker norms and more tolerance for dissent.” (Jackson et al. 2019; Gelfand et al., 2021). Some social challenges may be more successfully addressed by a tighter society and others by a looser society. More advanced natural language processing, such as Google’s BERT, Bidirectional Encoder Representations from Transformers (Devlin et al., 2019,) that captures more meaning-level information because it analyzes whole sentences at a time could also be used. A preliminary study by Liu et al. (2021) using BERT, showed that the prediction accuracy only decreased 4% when the words in the articles were scrambled into a random order. This suggests that the word vocabulary alone, rather than more sophisticated linguistic features, plays a significant role in differentiating lower-peace and from higher-peace countries. Our results here can also be tested by analyzing larger data sets that include more countries. The challenge here is to properly balance the increased data with the increased bias from the fewer news sources in countries where English is not the primary language.

The peace indices shown in Table 2, and their values normalized from 0-100 shown in Table 3, surprisingly, have different values for the same country. What sense can we make of those differences? Each index uses its own assumptions as to what are the indicators of peace and how to weight their relative importance. Could it be that they are all correct? Understandings of peace likely vary in different contexts. As noted by Roger Mac Ginty (2013), “Different communities are likely to define peace in different ways” (p. 59). We speculate that there is no one, single measure of peace. There could be different ways, ethnically, culturally, politically, socially, historically, economically, that countries can be peaceful and sustain their peaceful character. John M. Gottman and his collaborators (Coan and Gottman, 2007; Lisitsa, 2022) have identified the 4 most negative emotions that “describe communication styles that, according to our research, can predict the end of a relationship.” Those communication styles are: criticism, contempt, defensiveness, and stonewalling, which are all characterized by an underlying lack of emphatic connection. Do all lower-peace countries share this same underlying lack of empathic connections? In the opening sentence of Ann Karenina, Leo Tolstoy writes, “All happy families are alike; every unhappy family is unhappy in its own way.” Perhaps, peace may be just the opposite of Tolstoy’s families. Perhaps, there are many ways countries can be peaceful, but only one way that they are not peaceful.

## Data Availability

The data analyzed in the current study are available at:
https://drive.google.com/drive/folders/1jm7Z0YughSTdfk1T2zBrrF_xW7EANrMA?usp=sharing
A description of that data is available at:
https://github.com/mbmackenzie/power-of-peace-speech
The programs used to analyze that data in the current study are available at:
https://github.com/wpqc21/ArticleClassifier/tree/main/ArticleClassifierHSSC
https://github.com/smilelinnn/Article-Classification

## Author Contributions

All the authors contributed equally to this work.

## Competing Interests

The authors declare no competing interests.

## Ethics Statement

This article does not contain any studies with human or animal participants performed by any of the authors.

## References

A743 (2019) Which is Better? Systemic (Holistic) or Symptomatic (Reductionistic) Approach to Data Science, Oct 7, 2019. https://medium.com/@A743241/which-is-better-systemic-bottom-up-or-symptomatic-top-down-approach-to-data-science-9bae2afca518. Accessed 23 Jan 2023

Advanced Consortium of Cooperation, Conflict, and Complexity (2018) Sustaining Peace Project. http://sustainingpeaceproject.com. Accessed 2 Dec 2021

Bolívar A (2011) Language, Violent and Peaceful Uses of. In: DJ Christie (ed): The Encyclopedia of Peace Psychology. Blackwell Publishing Ltd, https://doi.org/10.1002/9780470672532.wbepp146

Caplan R (2019) Measuring Peace: Principles, Practices, and Politics. Oxford University Press, USA

Coan JA, Gottman JM (2007) The Specific Affect Coding System (SPAFF). In: Coan JA, Gottman JM (eds) Handbook of Emotion Elicitation and Assessment. Oxford University Press, New York, p 267-285

Coleman PT, Deutsch M (eds) (2012) The Psychological Components of Sustainable Peace. Springer, New York

Coleman, P.T., Fisher, J., Fry, D.P., Liebovitch, L. Chen-Carrel, A., Souillac, G. (2020). How to Live in Peace? Mapping the Science of Sustaining Peace: A Progress Report. American Psychologist

Cross-validation (statistics). Wikipedia. https://en.wikipedia.org/wiki/Cross Accessed 4 April 2023.

Devlin J, Chang MW, Lee K, Toutanova K (2019) BERT: Pre-training of Deep Bidirectional Transformers for Language Understanding. arXiv:1810.04805.

Deutsch M, Coleman PT (2016) The psychological components of a sustainable peace: An introduction. In Brauch HG, Spring UO, Grin J, Scheffran J (eds) Handbook on sustainability Transition and Sustainable Peace, Springer, New York, p. 139

Diehl PF, Goertz G, Gallegos Y. (2019) Peace data: Concept, measurement, patterns, and research agenda. Con Man Pea Sci. https://doi.org/10.1177/0738894219870288

Diehl PF (2016) Exploring peace: Looking beyond war and negative peace. Int St Qua, 60(1):1–10

Ezeibe C (2021) Hate Speech and Election Violence in Nigeria. J Asi Afr Stu, 56(4): 919–935. https://doi.org/10.1177/0021909620951208

Firchow P, Ginty RM (2017) Measuring peace: Comparability, commensurability, and complementarity using bottom-up indicators. Intl St Rev 19(1): 6-27. https://doi.org/10.1093/isr/vix001

Fisher J, Chen-Carrel A, He Y, Mei Q & Wideroth A (2020). Measuring Sustainable Peace: Assessing a new theoretical model through the application of existing peace data. Unpublished manuscript.

Friedrich P (2007) English for peace: Toward a framework of Peace Sociolinguistics. Wor Eng 26(1): 72–83. https://doi.org/10.1111/j.1467-971X.2007.00489.x

Friedrich P (2019) Applied Linguistics in the Real World. Routledge, London

Fry DP (2006) The human potential for peace: An anthropological challenge to assumptions about war and violence. Oxford University Press, USA

Fry DP, Souillac G, Liebovitch LS,, Coleman PT, Agan K, Nicholson-Cox E, Mason D, Gomez FP, Strauss S (2021). Societies within peace systems avoid war and build positive intergroup relationships. Humanities and Behavioral Sciences Communications 8, 17. https://doi.org/10.1057/s41599-020-00692-8.

Fund for Peace (2019) Fragile States Index Annual Report 2019. The Fund for Peace, Washington DC. https://fundforpeace.org/2019/04/10/fragile-states-index-2019/. Accessed 1 Aug 2022

Gelfand MJ, Jackson JC, Pan X, Nau D, Pieper D, Denison E, Dagher M, Van Lange PAM, Chiu CY, Wang M (2021) The relationship between cultural tightness–looseness and covid-19 cases and deaths: a global analysis. Lan Pla He 5(3):e135–e144

Goertz G, Diehl PF, Balas A (2016) The puzzle of peace: The evolution of peace in the international system. Oxford University Press, USA

Gomes de Matos F (2000) Harmonizing and humanizing political discourse: The contribution of peace linguists. Peace and Conflict: J Peace Psych 6:339-344

Helliwell JF, Layard R, Sachs JD (2019) World Happiness Report 2019. https://worldhappiness.report/ed/2019/. Accessed 1 Aug 2022

Institute for Economics & Peace (2019a) Positive Peace Report 2019: Analysing the Factors that Sustain Peace. http://visionofhumanity.org/reports. Accessed 1 Aug 2022

Institute for Economics & Peace (2019b) Global Peace Index 2019: Measuring Peace in a Complex World. http://visionofhumanity.org/reports. Accessed 1 Aug 2022

Institute for Economics & Peace (2021) Positive Peace Report 2021: Analysing the Factors that Sustain Peace. http://visionofhumanity.org/reports. Accessed 1 Aug 2022

Investopedia Team (2022)Top-Down vs. Bottom-Up: What's the Difference?, Updated September 06, 2022. https://www.investopedia.com/articles/investing/030116/topdown-vs-bottomup.asp Accessed 23 Jan 2023

Jackson JC, Gelfand MJ, De S, Fox A (2019) The loosening of american culture over 200 years is associated with a creativity– order trade-off. Nat Hum Beh 3(3):244–250

Jung J, Lee H, Kwon HJ, Mackenzie M, Lim TY (2021) power-of-peace-speech. https://github.com/mbmackenzie/ power-of-peace-speech/. Accessed 2 Dec 2021

Karlberg M (2011) Discourse Theory and Peace. In: Christie DJ (ed) The Encyclopedia of Peace Psychology. Blackwell Publishing Ltd, p. 87

Kimotho SG, Nyaga RN (2016) Digitized ethnic hate speech: Understanding effects of digital media hate speech on citizen journalism in Kenya. Adv Lan Lit Stu 7(3): 189-200

Lane H, Howard C, Hapke HM (2019) Natural Language Processing in Action. Manning, Shelter Island, NY

Leetaru K (2013)The GDELT project. https://www.gdeltproject.org/

Lisitsa E (2022) The Four Horsemen: Criticism, Contempt, Defensiveness, and Stonewalling. https://www.gottman.com/blog/the-four-horsemen-recognizing-criticism-contempt-defensiveness-and-stonewalling/. Accessed 29 Jul 2022

Liu H, Qi H, Wu X, Zhou Y, Zhu W. Power of Peace Speech https://github.com/wz2536/power-of-peace-speech_CapstoneFall2021 Accessed 2 Dec 2021

Luhmann N (1987) Soziale Systeme: Grundriß einer allgemeinen Theorie. Suhrkamp, Frankfurt

Mac Ginty R (2013) Indicators+: A proposal for everyday peace indicators. Eval Pro Pl 36(1): 56-63. https://doi.org/10.1016/j.evalprogplan.2012.07.001

Mahmoud Y, Makoond A (2017) Sustaining peace: What does it mean in practice? International Peace Institute

Ngabonziza AJD (2013) The Importance of Language Studies in Conflict Resolution. J Afr Con Pea Stu 2(1): 33-37. http://dx.doi.org/10.5038/2325-484X.2.1.4

NOW (2021) News on the web corpus. https://www.english-corpora.org/now/

PeaceTech Lab (2020) Combating Hate Speech. https://www.peacetechlab.org/hate-speech. Accessed 1 Aug 2022

Pedregosa F, Varoquaux G, Gramfort A, Michel V, Thirion B, Grisel O, Blondel M, Prettenhofer P, Weiss R, Dubourg V, Vanderplas J, Passos A, Cournapeau D, Brucher M, Perrot M, Duchesnay E (2011a). Scikit-learn: Machine Learning in Python sklearn.ensemble.RandomForestClassifier -. J Mach Learn Res, 12:2825–2830, 2011b.

Pedregosa F, Varoquaux G, Gramfort A, Michel V, Thirion B, Grisel O, Blondel M, Prettenhofer P, Weiss R, Dubourg V, Vanderplas J, Passos A, Cournapeau D, Brucher M, Perrot M, Duchesnay E (2011b) Scikit-learn: Machine learning in Python sklearn.linear model.logisticregression J Mach Learn Res 12: 2825–2830

Preacher KJ, Rucker DD, MacCallum RC, Nicewander WA (2005) Use of the Extreme Groups Approach: A Critical Reexamination and New Recommendations. Psy Met APA, 10(2): 178–192

Raschka S, Mirjalili V (2019) Python Machine Learning 3rd Ed. Packt>, Birmingham, UK
Reutter A (2020) Top-Down vs. Bottom-Up Approaches to Data Science, June 9, 2020. https://blog.dataiku.com/top-down-vs.-bottom-up-approaches-to-data-science. Accessed 23 Jan 2023
Soral W, Bilewicz M, Winiewski M (2018) Exposure to hate speech increases prejudice through desensitization. Agg Beh, 44(2): 136-146
United Nations Development Programme (2021) Human Development Index. https://hdr.undp.org/data-center/human-development-index#/indicies/HDI. Accessed 1 Aug 2022
Voukelatou V, Pappalardo L, Miliou I, Gabrielli L, Giannotti F (2020) Estimating countries' peace index through the lens of the world news as monitored by GDELT. Paper presented at IEEE 7th International Conference on Data Science and Advanced Analytics (DSAA) Sydney, 6-9 October 2020, p 216-225
Voukelatou V, Miliou I, Giannotti F, Pappalardo L (2022) Understanding peace through the world news. EPJ Data Science 11:2 https://doi.org/10.1140/epjds/s13688-022-00315-z
Yusuf S, Voss SJ (2018) The generations for peace institute compendium of participatory indicators of peace. Generations for Peace Institute

## Tables

Table 1 Countries studied.
Table 2 Peace indices from 5 sources. GPI=Global Peace Index, PPI=Positive Peace Index, WHI=World Happiness Index, FSI=Fragile States Index, and HDI=Human Development Index.
Table 3 Peace indices in Table 2 each independently, linearly scaled for these countries where 0 is lowest-peace and 100 is highest peace.
Table 4 Table 3 color coded independently for each index, lower-peace group=red, higher-peace group=green, and intermediate-peace group=yellow. Countries were identified as lower-peace if they had 3 or more peace indices in the lowest group, higher-peace if they had 3 or more indices in the highest group, and the remaining countries as intermediate-peace.
Table 5 Data for the countries of the 3-class training set, lower-peace=red, intermediate-peace=black, and higher-peace=green.
Table 6 Data for the countries of the 2-class training set, lower-peace=red and higher-peace=green.
Table 7 Performance measures over 20 runs of each machine learning method, each value is mean ± sem. 3-class models are: lower-peace, intermediate-peace, and higher-peace. 2-class models are: lower-peace and higher-peace.
Table 8 Machine learning (ML) peace index compared to the other peace indices. Training set of countries: lower-peace=red and higher-peace=green.

## Figure Legends

Figure 1. Strategy of the analysis.

Figure 2. The 100 most frequent words for the higher-peace and lower-peace countries. Yellow indicates the words of highest feature importance in making the higher-peace and lower-peace classification by the random forest feature importance method.

Figure 3. Word cloud of the words of highest feature importance, with their size scaled to their frequency of occurrence, for higher-peace countries.

Figure 4. Word cloud of the words of highest feature importance, with their size scaled to their frequency of occurrence, for lower-peace countries.